\documentclass[12pt]{article}   

\usepackage{graphicx}

\usepackage{mathrsfs,amsbsy,amssymb,latexsym,amsfonts,amsmath}

\usepackage{cite}

\begin{document}

\begin{titlepage}

\begin{flushright}
OIQP-11-11
\end{flushright}

\vspace{\baselineskip}

\begin{center}
\large \bf An Idea of New String Field Theory\\
-- Liberating Right and Left Movers --
\footnote{To appear in the Proceedings of the 14th  Workshop 
``What comes beyond Standard Models'', Bled July 11 - 21, 2011, 
eds. Norma Mankoc Borstnik, 
Holger B. Nielsn and Dragan Lukman.} 
\end{center}

\vspace{1.5\baselineskip}

\begin{center}
Holger B.\ Nielsen${}^{a}$ and Masao Ninomiya${}^{b}$
\end{center}

\vspace{\baselineskip}

\begin{center}
\it
${}^{a}$Niels Bohr Institute, University of Copenhagen,\\
17 Belgdamsvej, DK 2100 Denmark
\end{center}

\vspace{0.5\baselineskip}

\begin{center}
${}^{b}$Okayama Institute for Quantum Physicvs,\\
Kyo-yama 1-9-1  Kitaku, Okayama 700-0015, Japan
\end{center}

\vspace{2\baselineskip}
\begin{abstract}

We develop the idea for a new string field theory of ours that was proposed
earlier in a very rudimentary form in a talk in the Symposium of Tohwa 
University~\cite{one}. 
The main point is to describe the system of strings in the Universe by means
of the images of the $\tau$-derivatives of the right and left mover parts,
$\dot{X}^{\mu}_{R}$ and $\dot{X}^{\mu}_{L}$ respectively. The major progress 
since the Tohwa-talks ~\cite{one}
is to imagine a discretization of the $\tau_{R}=\tau-\sigma $, $\tau_{L}=\tau+\sigma $ variables for the right
and left movers respectively. We then observe that, by using $\dot{X}^{\mu}_{R,L}$
values at only the descritized even-numbered sites, we can set the commutation rules
for second quantization without any contradiction. In fact we can quantize the objects
described by these even numbered images. A light-cone frame description of the string
field theory in this way is presented. 
\end{abstract}


\end{titlepage}

\section{Introduction}\label{sec:intro}

There has been various proposals for how to make a second quantized string
theory~\cite{two, three, four, five}. It is meant a formalism in which it is 
possible to have a physical
space analogous to the Fock space in quantum field theory so as to describe
an arbitrary number of strings \cite{strings}.

It is the purpose of the present article to propose
(by taking up the line of our earlier Tohwa proceedings contribution 
\cite{one}  the basic ideas for 
a different type of string field theory than the usual one. 
It may be denoted by the code words that in our formalism right and left movers are 
liberated. It is clear that the present formalism described here is different from 
the earlier attempts by the fact that in our present scheme we do not distinguish as
many different (Fock space like) states as the other string field theories do.

In fact when two strings pass through the same point of the 25 dimensional space
at the same moment of time, one just have four pieces of strings with common
point as the end points.

But then it is a priori not obvious which pairs out of these four pieces are to be
considered belonging to the ``same string''. In the earlier string field theories
it is, so to speak, part of the physical degrees of freedom of the multi-string
state which parts of the strings make up one string and which part of the next string does. 
In our formalism two states including the same pieces of strings are considered 
the same state. That is to say, in our Fock space like formalism we do not 
distinguish the two states, which only deviate from each other 
with respect to which of the pieces 
of strings are on one string and which are 
on the  the next string.

In this sence our formalism or model has much fewer Fock space like states than the usual formalism.
However it should be noted that both our formalism and earlier string field
theories have at the end infinitely many states. Thus in this sense of just counting it does not make
much sense to say our formalism has fewer states: But it is clear that some of the states
in our formalism must correspond to several states in the earlier formalisms.

One might speculate that a reason for the relatively large degree of complication in the earlier
formalisms such as Kaku-Kikkawa and Witten, etc.~\cite{two, three, four, five},
is due to the fact that from the point of view of our formalism, these formalisms must carry 
around with an appreciable amount of superfluous information,i.e. superfluous degree of
freedom.

Precisely these extra degrees of freedom telling which pieces of string hang together 
making up string number 1, string number 2, and so on are physically what changes 
just in the very moment of a scattering. Indeed what must happen - in a classical way of
thinking - for locality reasons is that pieces of strings are shuffled around from one string 
to the other one in the infinitesimally short moment of time in which only locally 
interacting strings interact. At this infinitesimal time the places in 25 
dimensional space where some piece of string is present cannot change. There is not 
sufficient change for the positions or momentum densities of the pieces of string 
going on in the zero time alotted to it.

So the only thing that can truly happen during the basically zero length of time in which 
the strings are in touch and thus allowed to interact from the principle of 
locality, is that the various pieces of strings can be redistributed among the
string mumber 1, string number 2, etc. But as just described then this redistribution that 
can happen is in our own formulation only a thought property without physical content,
while in the older string field theory it is physical.

Here we have already alluded to the major point of our formalism that
nothing happens during the scattering. In a way scattering is in our formalism 
immaterial! 

Actually our string field theory makes nothing happen at all. In our formalism
we, do not describe the string themselves. Rather we make our formalism
to concern instead the right and left moving components of the string position 
vector in 25 + 1 dimentional space time, $X^{M}(\tau,\sigma )=X^{M}_{R}(\tau,\sigma) + X^{M}_{L}(\tau,\sigma)$. 
It is well known that considering closed strings - so that we can avoid here 
the end point reflections - the right mover $X^{M}_{R}(\tau-\sigma)$ and the left one $X^{M}_{L}(\tau+\sigma)$  
only depend respectively on $\tau_{R} = \tau - \sigma$ and $\tau_{L} = \tau + \sigma$.
That is to say that if we describe them then in terms of these variables -
$\tau_{R}$ and $\tau_{L}$ respectively thinking of them as replacements for $\sigma$ 
in the right mover $X^{\mu}_{R}(\tau-\sigma)$ and in the left mover 
$X^{\mu}_{L}(\tau+\sigma)$ not vary in any time anymore. Looking at it in this way 
a description using right and left movers instead of the usual $X^{\mu}(\tau,\sigma)$ 
leads to the possibility of no development at all: As time goes nothing happens 
provided we use the variables $\tau_{R}=\tau-\sigma$, $\tau_{L} = \tau+\sigma$ respectively 
for right and left moving part. During the very moments of scattering also nothing has time 
to happen in our point of view. Different way of concerning pieces of string 
organised to form string number 1, string number 2, etc. should not be counted 
as physical degrees of freedom (in our scheme).

Let us immediately conclude that, as we shall see below 
the fundamental formalism 
in our description does not have genuine time development. We could say that it is a kind of 
Heisenberg picture. However, it is important to notice that this does not mean that 
in practice one cannot describe scattering and calculate Veneziano amplitudes. 
Having prepared a state with typically several strings present one can translate 
such a state into Fock space like state in our formalism. Next we then ask what is 
the probability for finding a certain set of ``outgoing''  strings. To evaluate an answer 
for that sort of S-matrix question one now has to also translate the final state 
into a last state of the Heisenberg nature of our picture and further one 
formally calculates as if the S-matrix were the unit operator. The translations into 
our formalism of single string descriptions - which involves calculations of 
left and right movers and their quantum fluctuations - can turn out complicated enough 
that even the (trivial) overlap $\langle f|i \rangle$ can turn out to become the Vaneziano model.

In the following section 2, we start by describing how we think of the $\dot X^{\mu}_{R}$ and $\dot X^{\mu}_{L}$
images in a discretized way as consisting of chains of ``objects''.
In section 3, we shall report the theorem about classical string scattering 
which is absolutely crucial for our formalism. It is this theorem that guarantees 
the conservation of the images in 25 dimensional space-time of the maps 
$\dot X^{\mu}_{R}$ and $\dot X^{\mu}_{L}$ from all the strings present. In subsection 3.2 
we tell how our formalism is to be adjusted to having also open strings. In this case we have, 
contrary to the only closed string case, not two but only one image of $\dot X^{\mu}_{R,L}$ space.
In section 4 we review a bit string theory and point to some problems with
constructing our type of Fock space (for ``objects'').
In section 5 we mention the remaining reparametrization freedom of the $\tau_{R}$ and 
$\tau_{L}$ variables. In section 6 we 
discuss that because we only represent the even-numbered
objects in our formalism the ``odd'' ones have to be recovered from the conjugate variables
of the even ones. The second quantization which is the main point of a string field theory
is then presented in section 
7. In section 8 the lack of time dependence or the Heisenberg
picture nature of our formalism is explained. This constitutes a worry for our model being 
totally trivial but in section 
9 we suggest that in spite of that our model is presumably
able to provide the Veneziano model scattering amplitudes, and if so of course it would
have proven its right to be considered a string field theory. In section 
10 we mention some of the
technicalities still waiting before our formalism should be considered. 
In section 
11 we present conclusion 
and a bit of outlook.

\section{Our main point, the ``objects''}\label{sec:ourmainpoint}
It should be stressed that we in the present paper do \underline{not} as in other string 
field theories start from string creating or string annihilating operators.
Rather we first relate each string to an in principle infinite number of ``objects''
which in turn are related to the right and left mover fields $\dot X^{\mu}_{R}$ and
$\dot X^{\mu}_{L}$ on the string.

For a given string state the right and left mover $\tau$-derived fields
\begin{equation}
\dot X^{\mu}_{R}(\tau_{R})=\dot X^{\mu}_{R}(\tau-\sigma)
\end{equation}
and
\begin{equation}
\dot X^{\mu}_{L}(\tau_{L})=\dot X^{\mu}_{L}(\tau+\sigma)
\end{equation}
are because of the periodicity in $\sigma$ periodic functions of respectively 
$\tau_{R}=\tau-\sigma$ and $\tau_{L}=\tau+\sigma$ with the same period as in $\sigma$ for
$X^{\mu}(\sigma, \tau)$.

Since the constraint conditions as is well known or seen below are 
$(\dot X^{\mu}_{R})^{2}=0=(\dot X^{\mu}_{L})^{2}$ a period of $\dot X^{\mu}_{R}(\tau_{R})$
or of $\dot X^{\mu}_{L}(\tau_{L})$ is a closed circle on the light cone in $25+1$ dimensional Minkowski space time.
(We shall for closed strings think of two different Minkowski spaces, 
one for $R$ and one for $L$, while
for open strings only \underline{one} Minkowski space time).

Now we imagine a discretization of the parameters $\tau_{R}$ and $\tau_{L}$. 
I.e. we approximate them by some integers in some way, and we can distinguish 
points in the discretization chains
as being ``even'' or ``odd''. To each such discretization point we associate, what we call, an ``object''.
Thus we got to each string associated a right and left chain of ``objects'' on the light cones in
one or two $25+1$ dimensional Minkowski spaces. Instead of going for directly making an annihilation 
operator for a whole string our main idea is to make annihilation operators for the \underline{``objects''}
and then you may construct an annihilation operator for a whole string as the product of those for all the
associated ``objects''.

It should be noted that there are a couple of small technical modifications to the just mentioned 
construction of ``objects'':

1)We only make the Fock space to describe our theory from the ``even'' numbered ``objects''

2)We shall rather use integrals over small regions about the discretization points than 
$\dot X^{\mu}_{R}$ and $\dot X^{\mu}_{L}$ themselves as the ``positions'' for the ``objects''

Also have in mind that since there are ($\infty$) many objects per string the number of strings
is \underline{not} related to the number of ``objects''.

\section{Our basic theorem for classical string scattering}\label{sec:basic theorem}

This basic theorem stating, that the images of $\dot X ^{\mu}_{R}$ and $\dot X ^{\mu}_{L}$  
are totally conserved for a system of strings even if they scatter, made up the major part 
of the talk in Tohwa symposium~\cite{one}. The rest of the proceedings of Tohwa were, from our present point of view, 
a too complicated and not the right attempt to our present string field theory, although 
in a very basic sence the present article repeats the main ideas in the Tohwa proceedings.

Our theorem sounds: For classical dual closed strings the over all the strings present,
(enumerated by $ i=1, 2, \cdots , N $) united images of the maps $ X^{\mu}_{R}:\{\tau_{R}\}\to $
$25+1$  dimensional Minkowski space, i.e. $    \{ \dot X^{\mu}_{R}(\tau_{R})|\tau_{R}\in \mbox{``period interval''} \}$
is constant as a function of time $X^{0}$ even when scatterings occur, (but may be only up to 
a nul set). 

Here ``period interval'' means the interval over which the variable $\tau_{R}=\tau - \sigma$
runs around the string in question. Often one would use a notation 
$\tau_{R}\in \mbox{``period interval''}= \left[\tau-2\pi, \tau \right]$.

When we talk about a moment of time $X^{0}$ in this theorem the idea is to consider all these 
$(\sigma, \tau)$-combinations for the various strings for which the $X^{0}(\sigma, \tau)=X^{0}$ 
the for the moment characteristic time $X^{0}$ value. Typically this will mean that there is for
a fixed moment $X^{0}$ relations between $\sigma$ and $\tau$ 
being different for the different classical
strings. It is, however, possible that we could have chosen the specific gauge - if we wish so even for all the 
strings present - that $X^{0}(\sigma, \tau) = \tau$. In this gauge - which is rather pedagogical - 
we can simply consider $\tau$  being the time, so that considering the moment $X^{0}$ means 
cosidering $\tau = X^{0}$. 

That we in the theorem assume the strings to be ``dual'' means that we let their motion be described by the usual
string theory dynamics, say by the Nambu action. Since we as our main point talk about the right and left movers,
$X^{\mu}_{R}(\tau_{R})$ and $X^{\mu}_{L}(\tau_{L})$, we must of course in order to get to them choose some gauge -
in the usual way too - so that the equations of motion for $X^{\mu}(\sigma, \tau)$ simplifies to the massless Klein-Gordon form
\begin{equation}
(\frac{\partial^2}{\partial \tau^{2}}-\frac{\partial^2}{\partial \sigma^{2}})X^{\mu}(\sigma, \tau)=0
\end{equation}
or
\begin{equation}
(\frac{\partial}{\partial \tau}-\frac{\partial}{\partial \sigma})
(\frac{\partial}{\partial \tau}+\frac{\partial}{\partial \sigma})
X^{\mu}(\sigma, \tau)=0
\end{equation}
so that we have the general solution to $X^{\mu}(\sigma, \tau)$ for each separate string on the form
$X^{\mu}(\sigma, \tau)= X^{\mu}_{R}(\tau-\sigma)+X^{\mu}_{L}(\tau+\sigma)$.
These functions $X^{\mu}_{R}(\tau-\sigma)$ and $X^{\mu}_{L}(\tau+\sigma)$  are not completely 
uniquely determined from the here given definition in as far as, we of course could add a constant $(25+1)$-vector
to $X^{\mu}_{R}(\tau-\sigma)$ and subtract the same vector from $X^{\mu}_{L}(\tau+\sigma)$.
This is one of the ``technical'' reasons that we formulate the theorem using 
$\dot X^{\mu}_{R}(\tau-\sigma)=\frac{d}{d \tau}X^{\mu}_{R}(\tau-\sigma)$
rather than simply by using $X^{\mu}_{R}(\tau - \sigma)$ itself.
The constant vector which could be added and subtracted is namely differentiated away (to 0)
and thus this ambiguity mentioned does not influence $\dot X^{\mu}_{R}(\tau-\sigma)$, but
only $X^{\mu}_{R}(\tau-\sigma)$ itself.

That the strings are assumed ``classical'' means that we do not in the theorem like in usual
string theory invoke quantum mechanics, but consider a theory of ideal infinitely thin strings 
obeying the equations of motion derived from the Nambu-action in a classical mechanics way.

The concept ``image'' in the theorem simply is meant in the function theory sense of being the set 
of points in the target space (here the $25+1$ dimensional Minkowski space-time $M_{25+1}$) 
into which some $\tau_{R}$ or $\tau_{L}$ is mapped.
For string number $i$ say in the gauge with $\tau =time $ mentioned the ``image'' is the image for the 
function
$$
\dot X^{\mu}_{R}(\tau-\sigma):[0, 2\pi]\to M_{25+1}.
$$
This image is of course formally written
$$
\mbox{``image''}=\left\{\dot X^{\mu}(\tau_{R})|\tau_{R}\in \mbox{``period interval''} \right\} \subseteq M_{25+1}
$$
where
$$
\mbox{``period interval''}=[\tau-2\pi, \tau].
$$

As we are in the theorem working classically there is of course nothing special about just the 
$26=25+1$ dimensions required
by 
bosonic  string theory quantum mechanically. So that we talk about $25+1$ dimensions here
shall only be considered a pedagogical trick to guide the quantum string theorist to
see how our thinking is related to usual quantum string theory.

When we only formulated the theorem for $\dot X^{\mu}_{R}(\tau-\sigma)$ but not for 
$\dot X^{\mu}_{L}(\tau+\sigma)$ also, it were just for 
simplicity of formulation. Of course the completely analogous theorem holds for the left mover
$X^{\mu}_{L}(\tau+\sigma)$, i.e. for the united images for these left mover functions,
is also valid.

\subsection{The proof of the Theorem.}\label{subsec:proof}

The proof may be performed by writing down the right and left mover representations for 
the various strings number $i=1, 2, ..., N$ and then use that classically the scattering must 
take place by reshuffling of the various pieces of strings specified by the point at which
two scattering strings have a point in common. We shall, however, not go through such a 
calculation, which may be found in~\cite{one}.

It may, however, be easier to argue as we already essentially did in the introduction: 
classically infinitely thin strings will generically only meet and interact momentarily. 
So all the time, seen from some frame, except for a null-set the strings do not touch. 
During the major time intervals in between the strings are completely free. For closed strings 
- as we assume at first - e.g. $\dot X^{\mu}_{R}(\tau_{R})$ is a periodic function of $\tau_{R}$ 
with the period of that for $\sigma$ on the string. That implies that the image from each string 
under $\dot X^{\mu}_{R}(\tau_{R})=\dot X^{\mu}_{R}(\tau-\sigma)$ is actually unchanged. 
Varying $\tau$ just means shuffling $\sigma$ around the circle, i.e. period. This is true for 
all the strings string 1, string 2, $\cdots$ present at the time considered. 
Thus also the united image 
$I_{mR}$ is constant as long as time stays inside one of the intervals in which there are no interactions. 
But that were up to a nul-set all time, and during these intervals  
the  $\dot X^{\mu}_{R}(\tau_{R})$ and $\dot X ^{\mu}_{L}(\tau_{L})$ cannot change either. So we have proven the theorem that 
$I_{mR}$ and  $I_{mL}$ are totally independent of the cutting moment of time space-like surface. 
One should note that this theorem already means that there are two time conserved quantities 
$\dot X ^{\mu}_{R}(\tau_{R})$ and $\dot X ^{\mu}_{L}(\tau_{L})$ per point on the strings. Since this is essentially the dimension 
of the phase space of the string-system, so many conserved quantities mean that at least 
in the first crude estimate the many string system is a solvable mechanical system!

\subsection{Open string cases}\label{subsec:open}

Above we considered a string theory with only closed strings. The extension to one with also open strings is, 
however, quite easy, but interestingly turns out to mean that we have only one common mover image 
$I_{m}=I_{mR}\bigcup I_{mL}$ to consider. It is only this united mover-image which is conserved 
in the case of there also being open strings. 

Locally on the interior of the string there is 
no difference between an open and closed string and we can for both open and closed find the solution
of the form
\begin{equation}
X^{\mu}(\sigma,\tau)=X^{\mu}_{R}(\tau-\sigma)+X^{\mu}_{L}(\tau+\sigma).
\end{equation}
The part $X^{\mu}_{R}(\tau-\sigma)$ represents a pattern moving along the string ``to the right''
if we consider $\tau$ the time. But then when this pattern reaches the end of an open string 
this pattern a priori disappears because there is then no more any string on which to find just
that pattern of $X^{\mu}_{R}(\tau-\sigma)$. This would a priori spoil our theorem for open strings,
and indeed it does spoil it. However, we can very luckily make a very similar theorem working
also for open strings.

Indeed the point is that at the end points of the open strings we have the wellknown boundary
condition
\begin{equation}
0=\frac{\partial X^{\mu}(\sigma, \tau)}{\partial \sigma}
=\frac{\partial X^{\mu}_{R}(\tau- \sigma)}{\partial \sigma}
+\frac{\partial X^{\mu}_{L}(\tau+ \sigma)}{\partial \sigma}
=-\dot X^{\mu}_{R}(\tau-\sigma)+\dot X^{\mu}_{L}(\tau+\sigma)
\end{equation}

valid for the endpoint $\sigma$-values, say $\sigma = 0, \pi$. 

For example $\sigma = 0$ we simply obtain
\begin{equation}
\dot X^{\mu}_{R}(\tau)=\dot X^{\mu}_{L}(\tau).
\end{equation}
This equation must actually be true for all $\tau$ and thus the right mover function
$\dot X^{\mu}_{R}(\tau)$ and the left mover function $\dot X^{\mu}_{L}(\tau)$ do actually coincide.
It is now easy to see that as some pattern in $\dot X^{\mu}_{R}(\tau-\sigma)$ runs
into the end point an individual pattern just runs out of the same endpoint now as represented by 
the left mover $\dot X^{\mu}_{L}(\tau+\sigma)$ instead of by the right mover. But that means that the
total image of {\em both $\dot X^{\mu}_{R}(\tau-\sigma)$ and  $\dot X ^{\mu}_{L}(\tau+\sigma)$
becomes constant in time} just as in our theorem for the closed strings. Now it is just that you have to unite 
both the $\dot X^{\mu}_{R}(\tau_{R})$ images and the $\dot X^{\mu}_{L}(\tau_{L})$ images before you the set
$I_{mR}\bigcup I_{mL}$ which is conserved.

Of course when in the only closed string theory you have separate conservation in time of $I_{mR}$ and
$I_{mL}$ then also the union of these two sets will be conserved. In this sense you have a stronger
statement in the only closed string than in a model with also open strings. 

Concerning the question of whether the string theory also is essentially solvable due to our theorem 
in the open string theory one has to contemplate if the curve-structure total image 
$I_{mR}UI_{mL}$ has enough information to match the degrees of freedom of the strings.
In fact it turns that the open string theory is also solvable quite similarly to the closed string one.
In fact the crucial point is that there is in the union set $I_{mR}\bigcup I_{mL}$ for every little 
(infinitesimal) piece of string both a little piece corresponding to the right mover pattern and 
the left mover pattern in that piece. Theory we get that - ignoring the discussion of the time and longitudinal
dimensions, and only caring for the transverse dimensions - the development of the strings is totally - 
except for nullset - predicted by the conserved quantities, namely our $I_{mR}\bigcup I_{mL}$.

So in both cases, with and without open strings, we argue that string theory is essentially solvable.

\section{How to make a string field  theory.}\label{sec:howto}

With the understanding in mind that the whole state classically of a system 
of strings is describable
by the unification sets $I_{mR}$ and $I_{mL}$ (in the closed string case), we 
at first get to the idea
that a second quantized string theory should be given by a Fock space (system) in which
one can put some ``objects'' into two Minkowski-spaces $M_{25+1R}$ and $M_{25+1L}$ in  a way as to form 1-dimensional
curves. The idea then is that these curves of objects in $M_{25+1R}$ and $M_{25+1L}$ shall then be identified
with the unified images $I_{mR}$ and $I_{mL}$.(see figure 1, in section 11).

Remembering the usual string theory constraints
\begin{eqnarray}
(\dot X^{\mu}_{(\sigma,\tau)})^{2}+(X'^{\mu}(\sigma, \tau))^{2}&=&0~,
\nonumber \\
\dot X^{\mu}(\sigma,\tau)X'_{\mu}(\sigma,\tau)&=&0~,
\end{eqnarray}
which in terms of the right and left mover fields become
\begin{equation}
(\dot X^{\mu}_R(\tau_{R}))^{2}=0
\end{equation}
and
\begin{equation}
(\dot X^{\mu}_{L}(\tau_{L}))^{2}=0
\end{equation}
we see that the union of images $I_{mR}$ and $I_{mL}$ are contained in the light cones in the 
two Minkowski spaces, $M_{25+1R}$ and $M_{25+1L}$. Thus it will only be necessary to have the possibility
to put the ``objects'' that shall make up the $I_{mR}$ and $I_{mL}$ onto these light cones.

\subsection{More quantization.}\label{subsec:more}

With the crude starting idea of representing the united images $I_{mR}$ and $I_{mL}$ by putting 
1-dimensional curves of what we call ``objects'' onto the light cones in the two Minkowski
space-times we of course now want to translate the idea into a truly quantized string theory 
so as to obtain along these lines a true string field theory.

What we lack in the quantization compared to the Fock space describing ``objects'' forming chains
representing the $I_{mR}$ and $I_{mL}$ is to make use of $X^{\mu}_{R}(\tau_{R})$ and $X^{\mu}_{L}(\tau_{L})$
being operators, and thus of course having the correct commutation relations
\begin{equation}
\left[\dot X^{\mu}_{R}(\tau_{R}), \dot X_{R}(\tau'_{R})\right]= \delta'(\tau_{R}-\tau'_{R})2\pi \alpha' 
\end{equation}

Thinking now of the analogue of second quantization by means of a Fock space of a theory (quantum field theory)
constructed from the wish to form states with an arbitrary number of particles, we remember that the Fock-space construction
must be based on a system of single particle basis states. These basis states in turn are (typically)
eigenstates of a system of {\em commuting} operators. For example one can construct the usual quantum field theory
by three momentum eigenstates. Then the Fock-space basis vectors can be characterized by a number of particles
distributed into a set of such three momentum states
\begin{equation}
|\vec{p}_{1}, \vec{p}_{2}, \ldots, \vec{p}_{N}\rangle 
\end{equation}
In analogy to this Fock-space construction we shall now - for the purpose of our string field theory - 
construct a Fock-space based on states characterized by a distribution of a number of the ``objects'' 
- we talked about - into states described by points on the light cones in the two Minkowski spaces
$M_{25+1R}$ and $M_{25+1L}$. This would then mean that states, on the light cones, should be described as
eigenstates of the $\dot X^{\mu}_{R}(\tau_{R})$ or $\dot X^{\mu}_{L}(\tau_{L})$, now considered as operators.

\subsection{The commutation problem.}\label{subsec:commute}

But then we meet the problem, that since the $\dot X^{\mu}_{R}(\tau_{R})$ for different $\tau_{R}$
values do \underline{not} commute, so that we cannot just consider a single set of 26 operators commuting with themselves and
each other. If namely $\dot X^{\mu}_{R}(\tau_{R})$ and $\dot X^{\mu}_{R}(\tau_{R})$ do not commute they cannot be
used directly to construct a basis. The trick we shall use to overcome this problem is the following:

We first imagine that we discretize the variables $\tau_{R}=\tau-\sigma$ and $\tau_{L}=\tau+\sigma$ so that
we imagine $\tau_{R}$ and $\tau_{L}$ to take in some way or another only integer values. For example it could be
that $\tau_{R}$ and $\tau_{L}$ were taken to be a constant times integers. Really we shall take some more 
complicated rule for discretization. Then the important point of our trick 
is to {\em throw away the points of discretized
$\tau_{R}$ and $\tau_{L}$ values for which the integer is odd.}
We throw these points away in the sense that we do not associate
an ``object'' directly to the  ``odd'' points, but rather only to the even $\tau_{R}$ or $\tau_{L}$ points.

The crux of the matter then is that if we discretize the $\delta '$-function 
commutator 
$[\dot X^{\mu}_{R}(\tau_{R}), \dot X^{\nu}]=2 \pi \alpha' \eta ^{\mu \nu}\delta'(\tau_{R}-\tau'_{R})$
in the obvious way leads to that this commutator is only non-zero when the integers corresponding to
$\tau_{R}$ and $\tau'_{R}$ deviate from each other by $\pm{1}$. Thus if we only keep in our attempts to associate 
``objects'' directly by the $\tau_{R}$'s associated with the even integers we avoid the problem that the different relevant 
$\dot X^{\mu}_{R}(\tau_{R})$'s do not commute.

\section{Inclusion of reparametrization in $\tau_{R}$ and $\tau_{L}$}\label{sec:inclusion}

After having presented the crude idea of discretizing the right mover variable $\tau_{R}=\tau-\sigma$ and the
left mover variable $\tau_{L}=\tau+\sigma$ we should like to be a bit more specific taking into account that 
there is a left over part in Nambu action originally present reparametrization of the cordinates
$(\sigma, \tau)$, namely that we can even after the gauge choice still transform
\begin{eqnarray}
\tau_{R} &\to& \tau'_{R}=f(\tau_{R})~, \nonumber \\
\tau_{L} &\to& \tau'_{L}=g(\tau_{R})~.
\end{eqnarray}
Here the transformation functions $f$ and $g$ can be any pair of increasing functions consistent with the
periodicity conditions for $\tau_{R}$ and $\tau_{L}$. In litterature one often works with complexified
$\tau_{R}$ and $\tau_{L}$ under names $z$ and $\bar{z}$ and correspondingly f and g being analytic,
but we use the more ``physical'' $\tau_{R}$ and $\tau_{L}$ being real.

In principle we could fix this left over reparametrization and discretize in any way we like, but it is rather suggestive
to define as the variables to be represented as the position of the ``objects'' on the light cones in 
$M_{25+1R}$ and $M_{25+1L}$ not simply $\dot X^{\mu}_{R}(\tau_{R})$ and $\dot X^{\mu}_{L}(\tau_{L})$, but rather
\begin{equation}
\int_{{{\mathrm{REGION \, OF}}\atop{\mathrm{DISCRETIZED}}}\atop{\mathrm{POINT}}} \dot{X}^{\mu}_{R}(\tau_{R})d\tau_{R}~,
\end{equation}
and
\begin{equation}
\int_{{{\mathrm{REGION \, OF}}\atop{\mathrm{DISCRETIZED}}}\atop{\mathrm{POINT}}} \dot{X}^{\mu}_{L}(\tau_{L})d\tau_{L}~.
\end{equation}
That is to say: By any discretization there is a natural range covered by the $n$'th point $\tau_{R}(n)$
in the series of discrete points, say from the middle point between 
$\tau_{R}(n-1)$ and $\tau_{R}(n)$, i.e. $\frac{1}{2}(\tau_{R}(n-1)+\tau_{R}(n))$ to the middle point between 
$\tau_{R}(n)$ and $\tau_{R}(n+1)$ , i.e. $\frac{1}{2}(\tau_{R}(n)+\tau_{R}(n+1))$. To reduce the reparametrization dependence - 
but we do not at first remove it - we could then define our ``object'' - position $X^{\mu}_{R}(n)$ by the integral over
the ``covered region'' by the $\tau_{R}(n)$ included discrete point
\begin{equation}
J ^{\mu}_{R}(n) \widehat{=} \int^{\frac{1}{2}(\tau_{R}(n)+\tau_{R}(n+1))}
_{\frac{1}{2}(\tau_{R}(n)+\tau_{R}(n-1))}
\dot X^{\mu}_{R}(\tau_{R})d\tau_{R}~,
\end{equation}
and analogously for the left mover
\begin{equation}
J ^{\mu}_{L}(n)=\int^{\frac{1}{2}(\tau_{L}(n)+\tau_{L}(n+1))}
_{\frac{1}{2}(\tau_{L}(n)+\tau_{L}(n-1))}
\dot X^{\mu}_{L}(\tau_{L})d\tau_{L}~.
\end{equation}
By doing so we avoid some of the reparametrization dependence, so that the only reparametrization dependence
left comes from the reshufling of the separation points as $\frac{1}{2}(\tau_{R}(n)+\tau_{R}(n-1))$ 
and $\frac{1}{2}(\tau_{L}(n)+\tau_{L}(n-1))$ separating the small regions ``objects'' by the discrete points.

This remainig reparamentrization dependence could be suggestively gauge fixed by fixing one of the 
26-vector components of $X^{\mu}_{R}(n)$ and of $X^{\mu}_{L}(n)$. One could e.g. fix the time components 
$X^{0}_{L}(n)$ and $X^{0}_{R}(n)$ to some small constants, but it would be better and match better with more
usual string theory formalism to use infinite momentum frame. In infinite momentum frame coordinates 
for the $25+1$ dimensional target space time, we choose the coordinates $X^{+},X^{-},X^{1},X^{2},\ldots, X^{24}$.
The metric tensor $\eta _{\mu \nu}$ is taken in this infinite momentum frame as 
\begin{eqnarray}
\eta _{\mu \nu}&=&-\delta_{\mu \nu} \qquad \mbox{for} \quad \mu, \nu=1,2,\ldots,24
\nonumber \\
\eta _{+-}&=&2 
\end{eqnarray}
In this inifine momentum frame we suggestively fix $J ^{+}_{R}(n)$ and $J^{+}_{L}(n)$ to some small values as the 
parametrization.

If we only keep the even $n$ integrals $J^{\mu}_{R}(n)$ and $J^{\mu}_{L}(n)$ and replace as a sufficiently
good approximation for the light descretization the deltaprime function $\delta'(\tau_{R}-\tau'_{R})$
in the commutator for the dervatives $\dot X^{\mu}_{R}(\tau_{R})$ by a discretized approximation we can
arrange that all the kept even $n$ $J^{\mu}_{R}(n)$'s commute with each other. What does not commute is the 
$X^{\mu}_{R}(n)$ for say an even $n$ are the two nearest neighbors $X^{\mu}_{R}(n-1)$ and
$X^{\mu}_{R}(n+1)$ (which are ``odd'' and thus not directly included in the 
Fock space construction).

\section{On recovering the odd-$n$ $X^{\mu}_{R}(\tau_{R}(n))$'s}\label{sec:recover}

In order to achieve an effective commutation of the different
$J^{\mu}_{R}(n)
=\int^{\frac{1}{2}(\tau_{R}(n)+\tau_{R}(n+1))}_{\frac{1}{2}(\tau_{R}(n-1)+\tau_{R}(n))}\dot X^{\mu}_{R}(\tau_{R})d\tau_{R}$,
we threw out of consideration at first the $J^{\mu}_{R}(n)$ quantities for odd values 
of the integer $n$. Since we argued in the classical discussion in the beginning of this article that the full function
set $\dot X^{\mu}_{R}$ is needed to describe the strings and that it would thus not be satisfactory to leave out every
second point (after discretization), some way of recovering the $J^{\mu}_{R}(n)$-degrees of freedom for $n$ being odd is needed. 
The point of this recovery of the odd $n$ $J^{\mu}_{R}(n)$'s being some physical variables must of course be that
they are essentially the canonically conjugate variables to the even-$n$ $J^{\mu}_{R}(n)$'s. It has turned out 
that we have some technical details in order to fully realize such a correspondance between the odd $n$ 
$J^{\mu}_{R}(n)$-variables and the conjugate momenta to the even-$n$ $J^{\mu}_{R}(n)$'s. So let us in the 
present article be satisfied by establishing this interpretation of the odd $X^{\mu}_{R}(n)$-variables as being given 
by the conjugate momenta of even $X^{\mu}_{R}(n)$-variables in the crudest approximation of looking only locally
along a string, but ignoring the problems of restricting to $\dot X^{\mu}_{R}(\tau_{R})$ to be for each string a 
periodic function,
and also at first the problems of the constraint equations.
Under this simplifying ``approximation'' we see that we can simply construct 
an odd-$n$ $X^{\mu}_{R}(n)$ to be 
proportional to the difference between the conjugate momenta of the $J^{\mu}_{R}(n\pm 1)$ for the nearest 
neighbors in the discretized series (representing $\tau_{R}$). That is to say that if we denote by
\begin{equation}
\Pi _{\mu R}(m)
\end{equation}
the canonical conjugate of $X^{\mu}_{R}(m)$ then we see that we can use - up to an unimportant constants or quantities
commuting with the even-$m$ $X^{\mu}_{R}(m)$'s - that 
\begin{equation}
J ^{\mu}(n)\propto \Pi _{\mu}(n+1)-\Pi _{\mu}(n-1)~. 
\end{equation}

We can note that considering locally the situation along the string there is - in first approximation - 
a match of degrees of freedom: There are (in the limit of many discretized values per unit length in $\tau_{R}$)
approximately equally many even and odd numbers locally. Of course there are equally many conjugate variables
$\Pi_{\mu R}(m)$ (with $m$ even) as there are variables $X^{\mu}_{R}(m)$ to which they are conjugate. Thus
there in first approximation just the right number of conjugate variables to represent the odd-$n$ 
$J^{\mu}_{R}(n)$ quantities. To get the right commutation property approximating a derivative of a delta
that a $X^{\mu}_{R}(n)$ shall commute with all the other $X^{\mu}_{R}(m)$'s except for the two nearest neighbors 
$X^{\mu}_{R}(m \pm 1)$, we of course need the odd $X^{\mu}_{R}(m)$ to be a linear combination of the two 
neighboring conjugate variables. The sign must be so that it is the difference to simulate the derivative of
delta function $\delta' (\tau_{R}-\tau'_{R})$.

\section{Second Quantization}\label{sec:second}

Let us stress the main idea of our string field theory attempt - before going into the technical details
and problems of interpretation in reality - by mentioning creation and annihilation operators for the 
already earlier mentioned ``objects''. In fact whenever we have as suggested above a Fock space described theory
in which the states are described by how some ``objects'' are distributed in some set of variables, 
$X^{\mu}_{R}$ (we have here deliberately left out the enumeration of the ``object'' $n$)
we can, provided the objects are either bosons or fermions, construct for every possible value of the set of variables
$X^{\mu}_{R}$ a pair of a creation and an annihilation variable $a^{+}(X^{\mu}_{R})$ and $a(X^{\mu}_{R})$.
Of course the meaning is as usual that by acting with $a^{+}(X^{\mu}_{R})$ an ``object'' is put into the state for a single
``object'' in which the $X^{\mu}_{R}$-variable takes the value $X^{\mu}_{R}$ mentioned in the creation operator symbol
$a^{+}(X^{\mu}_{R})$. Similarly of course the annihilation operator $a(X^{\mu}_{R})$ removes one ``object''
with variables $X^{\mu}_{R}$ if there are any such ``objects''; if not it just gives zero.

Analogous to how one in quantum field theory can construct second quantized field operators $\phi (\vec{x})$ as 
Fourier-expanded in the annihilation operators $a(\vec p)$ we can in our formalism use Fourier transforms of
our $a(X^{\mu}_{R})$-operators to construct annihilation operators for ``objects''with given values of the
conjugate momenta $(\Pi_{\mu R})$. (For symplicity you should here rather think of a non-relativistic quantum 
field theory than a relativistic one with all the for our analogy at first complicating details of a 
Dirac sea, etc). Once we can Fourier transform to obtain creation $\phi^{+} (\Pi_{\mu R})$ and annihilation operators 
for ``objects'' with definite $\Pi_{\mu R^{-}}$ eigenvalues, we can easily extend that to make annihilation 
$a_{0}(X^{\mu}_{0R})$ and creation operators $a^{+}_{0}(X^{\mu}_{0R})$ for odd-``objects'' in the space of
``odd-object'' $X^{\mu}_{0R}$ into which we let the $X^{\mu}_{R}(m)$ for $m$ odd take their values.

At least if we have in mind some chain of the ``objects'' present, then we can corresponding to 
such a chain construct a series of odd-numbered ``objects'' by means of the neighboring object 
$\Pi^{\mu}_{R}$. This means that by creating by means of the creation operators $a^{+}_{e}(X^{\mu}_{eR})$
an appropriate superposition of states representing chains of ``even'' objects one gets corresponding to that also
a distribution for the odd-objects. Of course since ``odd'' and ``even'' $J ^{\mu}_{R}$'s like the 
$\dot X^{\mu}_{R}(\tau_{R})$'s, which they represent, do not commute and thus of course one cannot produce by the object 
creating operators $a^{+}_{e}(X^{\mu}_{R})$ a one string state in the Fock space with well defined values for the
$X^{\mu}_{R}$'s for both ``odd'' and ``even''. If one, however, is satisfied to make only either the even points
or odd points it should be possible. This is just analogous to that you can only create particles into states
in agreement with Heisenbergs uncertainty relations.

We shall of course have in the case of only closed string theories both have a Fockspace in which one can put in
``objects'' being interpreted as having ``positions'' given by the right mover $X^{\mu}_{R}$ and another Fock space
being interpreted as for left mover even ``objects''. The full Fock space for the total theory of only
closed strings shall then be the Cartesian product $H_{R}\otimes H_{L}$ of the left mover and right mover Fockspaces, 
where we denoted the Fockspace for ``even'' rightmover objects $H_{R}$ while that for leftmovers were denoted $H_{L}$.

If one considers a string theory with open strings also, the rightmover waves get converted into left mover waves on
the strings at the end  points. Thus our theorem only constitutes the conservation of the
$X^{\mu}_{R}$'s - or $\dot{X}^{\mu}_{R}$ image and the $X^{\mu}_{L}$'s or $\dot{X}^{\mu}_{L}$ image \underline{together}.
We must therefore in the open string theory put together ``objects'' for both left and right movers into a single
sort of ``objects'' - and still we only represent the ``objects'' in the formalism, which correspond to even
points on the $\tau_{R}$ and $\tau_{L}$ discretizations. - so as to only construct \underline{one} single
Fockspace for ``objects'' $H_{\mu}$. Here $H_{\mu}$ is a Fockspace with states constructed as states with a set of 
a mixture of right mover and left mover ``objects'' (leaving only ``even type'') are present. We do not 
even distinguish in our formalism for the open-string case between ``objects'' being right mover or left mover.
If we consider them - as we shall for the $\dot X^{\mu}_{R}, \dot X^{\mu}_{L}$ degrees of freedom;
contrary to Fermionic, Neveu-Schwarz-Ramond - say bosons we postulate even symmetry of the wave function 
(= Fockspace states) under the permutation of a rightmover ``object'' with a left mover one, and not only 
under permutation of right with right and of left with left.

\section{Lack of Time Development.}\label{sec:lack}

We have now above sketched how to build up a Fock space for describing a string theory, meaning that we have 
put forward the idea for a ``string field theory''. If should be stressed that this formalism idea of ours is
(in an abstract sense) a Heisenberg picture formalism. By this we mean that the Fock space state is not like 
in the Schr{\o}dinger picture to be developed in time. Really we have transformed the string theory so much
that we do not have any time in it any longer. Actually the reader will remember that it were absolutely 
crucial for our string field theory that the equation of motion for the field $X^{\mu}(\sigma,\tau)$ on the 
string were solved (essentially) by writing this field $X^{\mu}(\sigma, \tau)$ as a sum of a right mover
$X^{\mu}_{R}(\tau-\sigma)=X^{\mu}_{R}(\tau_{R})$ field and a left mover field 
$X^{\mu}_{L}(\tau+\sigma)=X^{\mu}_{L}(\tau_{L})$. That means, that we solved the equations of motion and went into 
a description using $X^{\mu}_{R}(\tau_{R})$ and $X^{\mu}_{L}(\tau _{L})$ which has not even the time $\tau$
for the internal string motion in it any more.

But now the true greatness of the observations leading to our formation were (above) that 
{\em even during the scattering processes} - considered classically - 
{\em the \underline{images} of the $\dot X^{\mu}_{R}(\tau)_{R}$ and $\ \dot X^{\mu}_{L}(\tau_{L})$
   remain unchanged} 
up to a nullset. That is to say even with scattering
there should (in the considered approximation) be no change in the description in terms of our Fock spaces).
So in our formalism - if taken as an ontological model - the scattering process is not in correspondance
with  anything physically existing. One may be worried that we in our formalism have thrown away too many
degrees of freedom, although it is in principle only a nulset comapared to the rest of the degrees of 
freedom - the ones on the pieces of the strings not hit in the scattering -. You remember also, that it is precisely
this ``nullset'' of information which in fact concerns how the different pieces of strings are glued together,
which distinguishes our string field theory from the earlier competing string field theories of Kaku-Kikkawa
and Witten. In the following section we shall discuss the problems of our string field theory.

\section{Some Problems and Explanations.}\label{sec:some}

We can indeed be worried by getting proposed a model string field theory, in which the scattering of strings is
only something, we think about, but which has no ontological content. 

Actually we shall, however, claim that this ``no true scattering'' in the ontological sense is indeed in fact 
o.k. The first argument suggesting that this type of model is indeed acceptable as a string field theory is to
point out that one can in principle - and we hope in a later publication show that it is quite easily actually
possible to do the calculations - obtain the Veneziano model scattering amplitudes from our string field theory.

In principle we have in our string field theory a Heisenburg picture formulation. If one wants to calculate 
an S-matrix in a Heisenberg picture model one must implement the initial state $|i \rangle $ for which we want
$\langle f|S|i\rangle$ as a $t \to -\infty $ or initial state. This means that if in the state $|i\rangle$ we have a number
of strings in various states, we at first think of $|i\rangle$ as being of the form
\begin{equation}
|i\rangle=|\{X^{\mu}_{i}(\sigma, \tau_{\mathrm{init}})\}\rangle
\end{equation}
if the string states were given as eigenstates of the $X^{\mu}_{i}(\sigma, \tau=\tau_{\mathrm{init}})$ for string number
$i$ in the initial state. Really an eigenstate of the $X^{\mu}(\sigma, \tau_{\mathrm{init}})$ is not realistic because
there will in  practice  rather be eigenstates of mass squares of the strings. That would mean states which
could be written as superpositions of eigenstates for $X^{\mu}(\sigma, \tau)$'s. The point is that we should write 
explicitely the zero-point fluctuation into the superposition coefficients. At least it requires a little bit of 
calculation to transcribe the practically interesting string states into eigenstates of the string position 
$X^{\mu}(\sigma, \tau)$ or what for our purpose is more important into eigenstates for $X^{\mu}_{R}(\tau-\sigma)$
and $X^{\mu}_{L}(\tau+\sigma)$; but here it must be kept in mind that $X^{\mu}_{R}(\tau-\sigma)$ e.g. do not 
commute with themselves for different $\sigma$-values (or $\tau_{R}-\sigma$ values). So we can first rewrite a given string
state - as e.g. a mass eigenstate for string - into eigenstates of the right and left mover $\dot X^{\mu}_{R}(\tau_{R})$
$X^{\mu}_{L}(\tau_{L})$ say {\em after we have discretized} and then use the proposed trick in this article that
{\em only keeps the even-numbered points of $\tau_{R}$ and $\tau_{L}$ in the discretization}.
 
The important point to stress it that one can by a moderate amount of calculation translate a given initial state 
$|i\rangle$ in forms of e.g. mass and momentum eigenstates for (isolated) strings into a linear combination of eigenstates of
the $\dot X^{\mu}_{R}(n)$ and $\dot X^{\mu}_{L}(n)$ for $n$ even, which represent $\dot X^{\mu}_{R}(\tau_{R})$ and
$\dot X^{\mu}_{L}(\tau_{L})$. Imagining that we have written the states of all the strings in the 
incomming state $|i\rangle$ into eigenstate of the even-numbered $\dot X^{\mu}_{R}(n)$ and $\dot X^{\mu}_{L}(n)$ we can then set up 
a Fock space state, in which there are present the  ``objects'' corresponding to just the strings present in the
state $|i\rangle$. The main point is that states given as string states for several strings such as $|i\rangle$ can be translated
into superpositions of states with given distributions of ``objects'' and thereby into a Fock space state in our model.
 
Analogously one can of course corresponding to a thinkable final state $|f\rangle$ described as some - well possibly different - 
number of strings, construct a state in our Fock space (or in closed strings  only case in the Cartesian product of our two
Fock spaces) for ``objects'', corresponging to $|f\rangle$.
 
If we denote the Fock space (or Cartesian product of two) states correspondance to $|i\rangle$ and $|f\rangle$ as
\begin{eqnarray}
|i\rangle &\to& |i\rangle_{\mathrm{Fock}} \nonumber \\
|f\rangle &\to& |f\rangle_{\mathrm{Fock}}
\end{eqnarray}
then the usual S-matrix element $\langle f|S|i\rangle$ becomes in our Heisenberg picture model
\begin{equation}
\langle f|S|i\rangle=\langle f_{\mathrm{Fock}}|i\rangle_{\mathrm{Fock}}.
\end{equation}
That is to say, that formally we use the unit operator as the S-matrix. As we have, however, already pointed out the rewriting
from the physical string language to our Fock space involves making explicit zero-point fluctuations of the strings and thus is 
sufficiently complicated, that the mathematical expressions, we shall finally obtain as function of the external momenta
(i.e. the momenta of the strings in the initial $|i\rangle$ and final state $|f\rangle$) will not be completely trivial. Rather it is indeed
not at all excluded that we should indeed, as physically expected, since we after all started with a string theory,
obtain a Veneziano model scattering amplitude for the S-matrix.

Indeed we can see some intuitive arguments, that an integral form of the type expressing the Veneziano models will appear
naturally in the evaluation of the overlap $\langle f_{\mathrm{Fock}}|i_{\mathrm{Fock}}\rangle$. 

In spite of the zero point fluctuations, we must include in the descriptions of
$|i_{\mathrm{Fock}}\rangle$ and $|f_{\mathrm{Fock}}\rangle$, there is still a correlation between
neighboring ``even'' ``objects'' or ($X^{\mu}_{R}(n)$ and $X^{\mu}_{L}(n)$), so that an overlap contribution in which the series
of ``objects'' lying along a pieces of string in one of the strings in $|i\rangle$, does go into a similar piece in the series in
one of the strings in the final state $|f\rangle$ will become a larger contribution (numerically) compared to a contribution in which
the piece of string in $|i\rangle$ is matched to several pieces of strings in the $|f\rangle$ strings. This favouring with respect to getting large
overlap contribution, when neighboring ``objects'' in the $|i\rangle$-state are analogously neighboring in the $|f\rangle$-state, is of course 
what corresponds to a continuous curve being transformed from $|i\rangle$ to $|f\rangle$. That is to say, that if we represent a chain or
curve ($\tau_{R}\in a$ period) of $\dot X^{\mu}_{R}(\tau_{R})$ fluctuating around  a classical curve, then the favoured overlap is the one 
in which the curves in $|i\rangle$ are transformed so 
in a way with as little splitting as possible into the $|f\rangle$ state. This kind of speculation suggests,
that we shall evaluate the overlap $\langle f_{\mathrm{Fock}}|i_{\mathrm{Fock}}\rangle$ (giving us the S-matrix element $\langle f|S|i\rangle$) as a 
perturbation series, in which the successive terms are classified according to the number of breaking points in the 
(classical representative) curves of ``objects'' (on the light cone in $25+1$ space time). That is to say the dominant
term should be the one with the lowest number of breaks; the next has more breaks and there are in principle an infinite
number of terms having more and more breakings.

By this consideration of using for calculating the main term is some S-matrix element
$\langle f|S|i\rangle=\langle f_{\mathrm{Fock}}|i_{\mathrm{Fock}}\rangle$ the contribution in which there has been made the lowest number of breakings
and rejoining the $\dot X^{\mu}_{R}(\tau_{R})$ and $\dot X^{\mu}_{L}(\tau_{L})$ curves from the $|i\rangle$ state to go into 
matching the curves in the $|f\rangle$-state. But now even for a minimal number of breakings of the 
$\dot X^{\mu}_{R^{-}}$ and $\dot X^{\mu}_{L^{-}}$ curves in order to go from $|i\rangle$ to $|f\rangle$ the needed breaking places on the chains of
``objects'' can be placed at different places in the chains. Since even in the speculated approximation
of the overlap contributions with fewest breaks dominating the different contributions just deviating by different places
on the chains of the breaking will contribute - in principle - in a comparable way and all have to be included in this
dominant approximation.

Our suggestion - which really has to be true - is that this summation over {\em different breaking places on the chains} 
but keeping the total number of breaking the same (namely the minimal number needed) is of course in the limit of the 
discretization being in infinitesimal steps truly an integral that shall turn out to be identifiable with the integrations 
in the Veneziano model for the scattering in question.

This suggestion is almost unavoidably true from the point of view that we have constructed our model from describing
physical strings and that is already well understood how these integrations of the Veneziano model are related 
either to true Minkowski space gluing and splitting of strings, or better to complex analytical continuation 
of such integrals of the points of splitting and joing.

In this article we have formulated the theory in Minkowski space-time as is the true physical situation,
but in formal string theory one often write instead the integrals over some $z=\tau+i \sigma$ and 
$\bar{z}=\tau -i \sigma$ rather than using the real valued $\tau_{R}=\tau - \sigma$ and $\tau_{L}= \tau+\sigma$
used above in the present article, but it is for the moment our belief that the differnce is just a technical
detail of a contour deformation of the integration. 

\section{Waiting Technicallities.}\label{sec:waittech}

We must admit that we have in the present only delivered the crude but crucial ideas for our string field theory, 
but we have not really yet put forward the detailed formalism, as is indeed required. Basically we have 
of course to go through in a string field theory (=a second quantized single string theory) the various
gauge choices, constraint, and anomaly problem one has in a single string theory.

We shall really go not through these problems of proper quantization in the present article, but rather postpone 
it to a later article or leave it for a reader to formulate our model in some cleverly chosen gauge etc.

Let us, however, here give an idea about what problems we have in mind in order to bring our ideas to 
provide a definite string field theory formalism:

1) To make the formalism definite one has either to choose the reparametrization of $(\sigma, \tau)$ inside the 
strings completely and even to discretize in a definite way so that our $X^{\mu}_{R}(n)$'s and 
$X^{\mu}_{L}(n)$'s obtain a well defined meaning (rather than being gauge-monsters) or one has to formulate a 
remaining part of the gauge freedom is propagated into our ``object'' and later Fock-space formulation.
By this is meant that if we do not choose the gause (meaning reparametrization of $(\sigma, \tau)$) to the end
we must a gauge freedom surviving into the Fock space formulation.

One solution to this problem which we are working on consists in using infinite momentum frame and using the 
requirement that
\begin{equation}
X^{+}_{R}(n)=E
\label{eq:gc}
\end{equation}
where $E$ is a certain constant to fix together the discretization and the reparametrization of the type
\begin{equation}
\tau_{R}\to \hat{\tau}_{R}(\tau_{R})
\label{eq:^}
\end{equation}
where $\hat{\tau}_{R}$ is any increasing function of $\tau_{R}$. Having assumed (\ref{eq:gc}) we can namely easily say that
we choose $\tau_{R}$ to be proportional to the number in the chain of the ``objects''. In this way then the gauge
of the rudimentary gauge freedom (\ref{eq:^}) gets chosen.

If we now choose then the constraints equation
\begin{eqnarray}
0&=&X^{\mu}_{R}(n)X^{\nu}_{R}(n)g_{\mu \nu}  \nonumber \\ 
&=&
2X^{+}_{R}(n)X^{-}(n)- \vec{X}_{RT}(n)\vec{X}_{RT}(n)
\end{eqnarray}
which restricts $X^{\mu}_{R}(n)$ to lay on the light-cone in the $25+1$ dimensional (formal) Minkowski space-time, 
gets further restricted by and the $X^{\mu}_{R}(n)$'s turn out to lay on a parabola on this light cone only.
That means that only the $24$ ``transverse'' coordinates $\vec{X}_{RT}(n)$ of $X^{\mu}_{R}$, meaning 
the ones perpendicular to both the + and - coordinates, are independent variables since the + coordinate
$X^{+}_{R}(n)=E$ by and the $X^{-}_{R}(n)$ then becomes the one given from the constraints,
\begin{equation}
X^{-}_{R}(n)=\frac{{\vec{X}}_{RT}(n)\cdot {\vec{X}}_{RT}(n)}{2E}.
\end{equation}

To make these special constructions for $X^{-}_{R}(n)$ and $X^{+}_{R}(n)$ provides a problem - which would like to 
postpone to our later publication - because there is no obvious generalization of how we construct the odd
$X^{\mu}_{R}(n \ \mbox{odd})$ in terms of the conjugate variables to the even 
$n$, $\vec{X}_{RT}(n \ \mbox{even})$ neighbors for special + and - components.

For the transverse coordinates of the odd $n$, $X^{\mu}_{R}(n)$'s, we have to put for $n$
odd
\begin{equation}
\vec{X}_{R}(n)\propto \vec{\Pi}_{R}(n+1)-\vec{\Pi}(n-1)
\end{equation}
in order to have the commutators between the neighboring ``objects'' $X^{\mu}_{R}(n)$ corresponding to
$\delta'(\tau_{R}-\tau'_{R})$ proportional commutator required for the 
$\dot X^{\mu}_{R}(\tau_{R})$'s.

The same commutation rule can of course not be arranged for the $X^{-}_{R}(n)$'s if we let them be 
just fixed numbers for both odd and even $n$ and for $X^{-}_{R}(n)$'s being quadratic expressions in
the transverse degrees of freedom also the commutators will be a priori more complicated.
So to organize a consistent picture involving also the + and - components is postponed to a later article.

\subsection{Status of the Odd ``Objects''.}\label{subsec:status}
Let us towards the end stress that in our formalism where the Fock space(s) describe directly only the                       
\underline{even} ``objects'' the \underline{odd} ones have to be constructed 
- formally - from the differences of the conjugate momenta
$\vec{\Pi}_{RT}(n+1)-\vec{\Pi}_{RT}(n-1)$. But this means that if the order in which we imagine the even 
``objects''are organized into chains corresponding to strings get changed - by scattering - 
then it is even so that the from the even objects constructed odd ones are not only put into a different ordering,
but there are even some odd ones being replaced by different ones. 
If a piece of a chain of even ``objects'' is attached to another piece than before then a new odd object
that were not present before is being included.
But it is only of the order of a few odd ``objects'' that get replaced, while the numbers of both odd and even 
objects in total in a number of strings goes to infinite in the limit of discretization being very fine.
In this sense the number of replaced odd objects can be considered a null-set and thus in some sense
be negligible.

\section{Coclusion and Outlook.}\label{sec:conslusion}
\begin{figure}[h]
\begin{center}
\includegraphics[height=0.4\textheight]{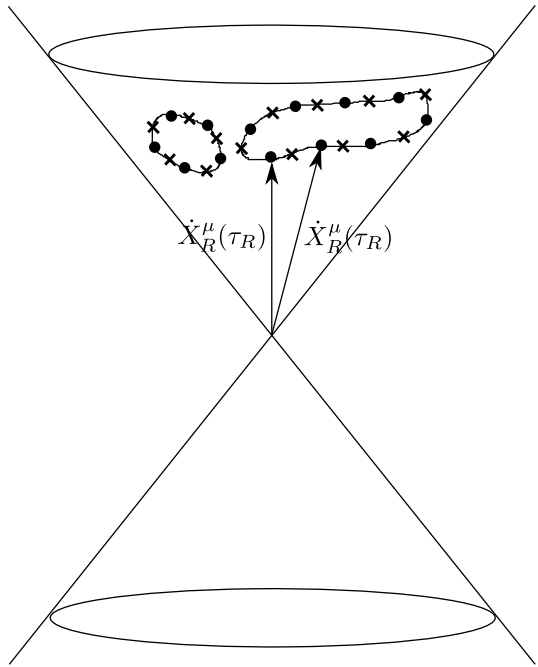}
\end{center}
\caption{Symbolic drawing of (one of) the $25+1$ Minkowski spaces
with the light cone, 
where the 25 +1 axes are $ \dot{X^0}, \dot{X^1}, \dot{X^2}, etc.$,
and on the latter a couple of closed
curves being the images of a couple of strings.
The crosses and dots alternating along these curves
symbolize the ``positions'' of the respectively
``even'' and ``odd'' ``objects''. The shown two curves correspond to that 
there are two strings represneted in the example drawn. }
\label{fig:1}
\end{figure}

We have put forward some ideas for constructing a string field theory, in the sense of a second quantized
theory of strings. The basic idea is 
{\em not to simply construct a Fock space in which one has states with various numbers of strings} 
such as one would at first think were to be done, and such as the already known theories, Kaku  Kikkawa or
Witten, have it. Rather $\underline{we}$ do not use the strings themselves but rather formally split them up 
into right mover and left mover degrees of freedom, mainly represented by $\dot X^{\mu}_{R}(\tau_{R})$
and $\dot X^{\mu}_{L}(\tau_{L})$. These separated degrees of freedom are then described by means of 
one-demensional chains of point constituents (after descretization) called ``objects''
(rather than for the strings themselves). It is for these ``objects'' 
we construct a Fock space. 
See Figure~\ref{fig:1} for an illustration of how the ``objects'' sit in one of the (two or one) Minkowski spaces.
Well, for the case of only closed strings we actually 
use {\em two} Fock spaces, one for the ``objects'' related to the right movers 
$\dot X^{\mu}_{R}(\tau_{R})$ and one to the $\dot X^{\mu}_{L}(\tau_{L})$. Since each string is described by
a large 
variable number of objects there is no (immediate) relation between the
number of ``objects'' in our Fock space(s) and the number of strings. Rather the strings may split and unite
without change in our Fock space(s). So it is in principle o.k. and can still allow for string scatterings
even if there is no time development of our Fock space states - for ``objects'' - at all. Indeed we 
initiated this article by arguing for a theorem for strings in classical approximation: For a theory with only closed strings 
the images under the mappings 
\begin{equation}
\dot X^{\mu}_{R}:~\bigcup_{i} \{\tau_{R}\ \mbox{range for string i}\} 
\longrightarrow M^{(R)}_{25+1}\mbox{(a Minkowski space time)}
\end{equation}
and
\begin{equation}
\dot X^{\mu}_{L}:~\bigcup_{i}\{\tau_{L}\ \mbox{range for string i}\} 
\longrightarrow M^{(L)}_{25+1}\mbox{(another Minkowski space time)}
\end{equation}
are conserved in time. This theorem implies a huge number of conservation laws, apart from null-sets the same number of 
degrees of freedom as the strings themselves. In this sense it means that 
up to the null-sets mentioned classical string theory,
even with scattering (splitting and unification of strings), is a solvable theory. In the language if representing the
strings by these images,
\begin{eqnarray}
I_{R}=\bigcup^{n}_{i=1} \left\{ \dot X^{\mu}_{R}(\tau_{R}) \right| 
                                \left. \tau_{R}\in \ \mbox {range for string $i$} \right\}\\
I_{L}=\bigcup_{i=1}^{n} \left\{\dot X^{\mu}_{L}(\tau_{L}) \right|
                             \left. \tau_{L}\in \ \mbox {range for string $i$} \right\}
\end{eqnarray}
there is no (time) development at all. This means that the string theory - even of several splitting and uniting strings
- has been $\underline{trivialized}$. Our string field theory formulated in terms of our ``objects'' means that we make used of 
this trivialized formulation of (classical) string theory. It is therefore our ``objects'' and thus also the Fock spaces
describing them do not develop at all.

One might therefore feel tempted to believe that we cannot obtain a scattering amplitudes - meaning the Veneziano model - 
in our scheme, since if scattering is not revealed by any change in the Fock spaces, how could there be any scattering 
amplitude? Surprisingly enough we hope, however, to have given suggestive arguments that indeed one can rather easily obtain
Venesiano scattering amplitudes from our formalism!

In some sense this surprise is related to the neglected ``null sets''. Such nul sets may in fact be thought of as telling about
how pieces of strings are glued together.

Whenever an applier of our formalism would ask for a scattering amplitude he would have calculated a 
not so simple
wave function in our Fock space for ``objects'' for the set of strings in the initial state $|i\rangle$. Similarly he must calculate a
wave function for the system of outgoing strings for which he asks the S-matrix element. This S-matrix element is then computed as
the overlap of the two constructed Fock space states. I.e. they are basically calculated as if the ``fundamental'' S-matrix
(in our Fock space) were just the unit operator. We gave arguments for series expansion of the in this way constructed
S-matrix leading easily to integrals very similar to the way that Veneziano models
are expressed.

Some of the major technical points or ideas put forward in the present article were to make our Fock space for the
``objects'' {\em only for the in a discretization of $\tau_{R}$ (or $\tau_{L}$) even numbered points}.

Another detail is that in the formulation of the $\dot X^{\mu}_{L}$ and $\dot X^{\mu}_{L}$ the usual consequence of the Virasoro - 
algebra restrictions becomes - classically at least - simply that our ``objects'' must lay on the light cone in the spaces 
$M^{(R)}_{25+1}$ and $M^{(L)}_{25+1}$ respectively.

Yet a suggested technical detail were to use rather than the $\dot X^{\mu}_{R}$ and $\dot X^{\mu}_{L}$ themselves to connect
with the ``objects'' these quantities integrated up over a small (discretization scale) interval in $\tau_{R}$ 
respectively $\tau_{L}$,
\begin{equation}
J^{\mu}_{R}(n)=\int^{\tau_{R}(n)+
\frac{\Delta \tau_R}{2}}_{\tau_{R}(n)-\frac{\Delta \tau_R}{2}} \dot X^{\mu}_{R}(\tau_{R}d \tau_{R})
\end{equation}
and
\begin{equation}
J^{\mu}_{L}(n)=\int^{\tau_{L}(n)+\frac{\Delta \tau_L}{2}}_{\tau_{L}(n)
-\frac{\Delta \tau_L}{2}} \dot X^{\mu}_{L}(\tau_{L}d \tau_{L})
\end{equation}
(Here $\Delta \tau_R$ and $\Delta \tau_L$ are disretization intervals.)
Then reparametrization symmetry in $\tau_{R}$ respectively $\tau_{L}$ can be gauge fixed in say an inifinite momentum frame by putting the
$X^{+}_{R,L}=$constant. This would bring the $X^{\mu}_{R,L}$'s to be only on a certain ``parabola'' lying on the light cone.

We hope in forthcomming paper to a) Specify how by construction we achieve 
$(x^{\mu}_{R}(n))^{2}=0$ and thereby deduce the mass spectrum for a closed chain of ``objects'' and thus of a string.
b) Compelete the evaluation of the Veneziano scattering amplitude c) Investigate if we really obtain unitarity corrections to 
Veneziano amplitudes by evaluating the overlaps contributions in which there are more breaks in the correspondence between the 
incomming and the final state chains of ``objects'' used to evalutate the overlap $\langle f|i \rangle$, that
should give the S-matrix element.

\section*{Acknowledgement}\label{sec:Ack}
The authors acknowledge K. Murakami for informing them many relevant references. M.N. would like to 
thank the Niels Bohr Institute for hospitality extended to him during his stay there. H.B.N. thanks for room and
allowance to stay at NBI as an emeritus professor to the Niels Bohr Institute.
One of us (M.N.) is supported by the JSPS Grant-in-Aid for Scientific Research 
Nos. 21540290 and 23540332.

\end{document}